# A Cooperative Game Approach for Energy Management of Interconnected Microgrids

Mohadese Movahednia, Hamid Karimi, Shahram Jadid

*Abstract*— In this study, a cooperative game model is presented to schedule the day-ahead operation of multi-microgrid (MMG) systems. In the proposed model, microgrids are scheduled to achieve a global optimum for the cost of the multi-microgrid system. The minimum cost is achieved by transactions of microgrids with each other. Also, price-based demand response is implemented in the model to build a cost-reducing opportunity for consumers. Applying Shapley value, the optimum cost of the MMG system is fairly allocated between microgrids. To enhance the confidence level of results, data uncertainties are incorporated into the model. The uncertainties of renewable outputs, demand, and prices of trading with the main grid are applied in the model. The presented model is developed as a mixed-integer linear programming problem, and its efficiency is evaluated on a standard test system containing three microgrids. The cost of the MMG system when microgrids form a cooperative game is compared with the isolated status that microgrids do not transact energy with each other. The results indicate that the cost of the MMG system is declined using the proposed cooperative model in comparison with the isolated mode. Also, the cost of microgrid 1, microgrid 2, and microgrid 3 are improved by 2.4, 2.7, and 11.8%, respectively. Therefore, all the microgrids have an incentive to participate in the cooperative game, and both the total cost and each microgrid cost are improved in the cooperative game.

## I. Nomenclature

**Sets**

| | |
|---|---|
| $t$ | sets of time intervals |
| $k$ | sets of microgrids |
| $i$ | sets of days |

**Abbreviations**

| | |
|---|---|
| MG | microgrid |
| MMG | multi-microgrid |
| CDG | controllable distributed generator |
| BESS | battery energy storage system |
| PV | photovoltaic |
| EMS | energy management system |
| DR | demand response |

**Constants**

| | |
|---|---|
| $P_{min}$ | min production limits of CDG |
| $P_{max}$ | max production limits of CDG |
| $\lambda^{buy}_{(t)}, \lambda^{sell}_{(t)}$ | per-unit buying/ selling price of power from/ to the main grid at $t$ |
| $\lambda^{cdg}(t)$ | per-unit cost of CDG generation |
| $P_{load(t)}$ | demand before implementing demand response at $t$ |
| $P^{L-con}_{(t)}$ | controllable demand at $t$ |
| $P^{l-fix}_{(t)}$ | base demand at $t$ |
| $P^{pv}_{(t)}$ | output generation of PV at $t$ |
| $OF^{max}_{(t)}, IF^{max}_{(t)}$ | highest outflow and inflow of demand from/ to period $t$ |
| $P^{cap}_B$ | BESS capacity |
| $\eta^{BTB}_B$ | converters efficiency connected to BESS |
| $L^{B-}$ | discharging losses of BESS |
| $L^{B+}$ | charging losses of BESS |
| $\delta_B$ | BESS self-discharge rate |

**Variables**

| | |
|---|---|
| $P^{cdg}_{(t)}$ | CDG output generation at $t$ |
| $P^{short}_{(t)}$ | shortage power at $t$ |
| $P^{sur}_{(t)}$ | surplus power at $t$ |
| $P^{l-adj}_{(t)}$ | demand after demand response at $t$ |
| $P^{shift}_{(t,t')}$ | shifted demand from $t$ to $t'$ |
| $P^{B-}_{(t)}$ | BESS discharging power at $t$ |
| $P^{B+}_{(t)}$ | BESS charging power at $t$ |
| $P_{buy(t)}$ | buying power at $t$ |
| $P_{sell(t)}$ | selling power at $t$ |
| $SOC^B_{(t)}$ | state of charge for BESS at $t$ |
| $SOC^{B'}_t$ | state of charge of BESS at $t$ before considering self-discharge |

## II. Introduction

CLUSTERING distribution systems into several microgrids enhance the flexibility of the system. Microgrids (MG) are low voltage networks that consist of interconnected loads and distributed energy resources. A group of interconnected microgrids is called a multi-microgrid (MMG) system. The control and management of these large systems have become a major challenge in recent studies [1]. Multiple studies have been accomplished on the energy scheduling of multi-microgrid systems. Authors in [2] and [3] suggest a centralized energy management scheme to manage the MMG systems. In the proposed models, all the microgrids are managed by a central EMS. Microgrids are scheduled to provide service to the total system. Therefore, the cost of the multi-microgrid system is minimized constructively. This method is extremely useful for MMG systems with an individual owner. It is worth mentioning that this feature can be problematic in the case of different owners for the total system and MGs. Furthermore, the calculation burden on the central EMS will increase exponentially with the increase of MG numbers. Therefore, it is more practical for small systems. To decrease the computational complexity, decentralized methods are suggested in [4], [5], [6], and [7]. In decentralized methods, the main problem is decomposed into smaller independent subproblems which are solved by local EMSs. However, the global optimum is not guaranteed as the local EMSs aim to find their own optimal without the information from other EMSs. A hierarchical energy management model to manage multi-microgrid systems is presented in [8], [9], [10], and [11]. In the hierarchical strategies, each MG optimizes its costs and sends the results to the central Energy Management System (EMS). The central EMS runs a global optimization and informs MGs about the results. Finally, each MG improves its scheduling based on the received data from the central EMS and informs its units about their commitments. In the hierarchical strategies, each MG have the ability to schedule its units based on its own interests. This feature may lead to a greater total cost for the MMG system. However, it is a good choice for systems where MGs have different owners or priorities. Furthermore, the calculation burden on the central EMS is decreased effectively. However, it is worth mentioning that this method needs a complex structure.

In contrast with the above-mentioned methods, which have a conflict between minimizing the MMG system cost and minimizing each microgrid cost, using a cooperative game approach can be an effective method to minimize MMG system cost and each microgrid cost simultaneously. Cooperative games aim to achieve a global optimum point where no player has any incentive to leave the game, as they cannot make any further improvement in their interests. The cooperative game is first created to obtain the global optimum for all the players, and then the total gain or cost is fairly allocated between all the players. Therefore, this approach can be used for the energy management of MMG systems. In other words, a cooperative game can be developed for the MMG system where each microgrid is a player. In the developed game, the cost of the multi-microgrid system is optimized, and then the total MMG cost is fairly allocated between microgrids. Few studies have been done on the application of cooperative games in MMG system management.

In [12], [13], and [14], a cooperative game model is presented for day-ahead scheduling of MMG systems. The proposed models are deterministic approaches, and the uncertainty of data is neglected. In [15], a cooperative game to schedule multi-hub systems is suggested, and the Shapley value is used to allocate the total gain between energy hubs. The planning schedule of the MMG system is formulated in [16] and [17] as a cooperative model where the fair cost model contains operational costs and investment costs. In this paper, a cooperative game model is suggested to manage the MMG system where the total cost of the MMG system is minimized, and Shapley value is applied to fairly allocate microgrid costs.

Conventionally, the forecasted information of uncertain data is used in the optimization algorithms, and for the sake of simplicity, deterministic models are suggested. However, incorporating the uncertainty of data in the model increase the confidence level of results. In this study, the uncertainties related to the demand, renewable outputs, and prices of trading power with the main grid are considered in the suggested method. Different methods have been suggested to apply these uncertainties in the optimization algorithms. In [18], the worst case of load and renewable resources has been used. In [19], [20], and [21], a robust strategy has been used to consider the uncertainty of load and renewable outputs. Authors in [22] have suggested using Monte Carlo simulation to create stochastic variable scenarios to implement the uncertainty of load, sustainable outputs, and electricity per-unit costs. In [23], the estimated variance of the electricity per-unit costs has been utilized to consider the uncertainty of prices. In this study, the worst case of photovoltaic generation and demand predictions has been deliberated in the model. Also, the estimated variance of the electricity per-unit costs is used to apply the predicted price uncertainty in the model.

Demand response can be applied to flat the demand profile of microgrids and make cost-saving possible for the customers. Various methods such as incentive-based DR program, time-of-use pricing, direct demand control, critical peak pricing programs, to name but a few, are suggested in the literature [24], [25], [26]. Ahmadi et al. [27] proposed a non-cooperative bi-level optimization framework for energy management of the islanded MMG system. The authors utilized the DR programs to cover the uncertainty of renewable generation. However, the efficiency of the cooperation among MGs was not studied. A tri-level optimization framework had been suggested in [28] to evaluate the role of DR programs on the coordination between the MMG system and distribution system. Nevertheless, the probabilistic behavior of loads and renewable generation had been ignored. The role of DR programs on the reconfiguration of microgrid-based systems is the main achievement of [29]. Although renewable generation was integrated into the model, the role of battery energy storage systems was not evaluated. The flexible economic operation scheduling of the MMG system was modeled in [30] by the leader multi-follower framework. The authors utilized the shiftable DR programs to provide the opportunity for cost-saving for the MGs. However, the uncertainty of loads and renewable generation was not applied to the model. Also, the efficiency of the cooperation among MGs and energy storage system had been ignored. In this paper, a price-based DR is deliberated in the presented method.

The main contributions of this study are as follows:
- Energy scheduling of multi-microgrid systems is modeled as a cooperative game approach.
- Shapley value is applied to fairly allocate the cost of the MMG system between microgrids.
- Transactions of energy between microgrids has enhanced the flexibility of the MMG system and reduced the MMG cost.
- To increase the confidence level of results, the uncertainties related to the demand, renewable outputs, and

prices of trading power with the main grid are considered in the suggested method.
- A price-based DR is deliberated in the method to facilitate the cost-saving of customers.

The remainder of this paper is structured as follows: The configuration of the suggested system and concepts, including the DR program and uncertainties, are presented in section III. The proposed cooperative game to determine the day-ahead schedule of units in the MMG systems and fairly allocate the cost of microgrids is presented in section IV. The simulation outcomes and conclusion are provided in sections V and VI, respectively.

## III. SYSTEM MODEL

A test system with three microgrids, which is depicted in Fig.1, is used to analyze the performance of the proposed model. Noteworthy, the proposed method is a general model and is not affected by the MG numbers. Each MG includes load, a Controllable Distributed Generator (CDG), Photovoltaic (PV), and Battery Energy Storage System (BESS). The Energy Management system (EMS) is responsible for energy scheduling for all the microgrids.

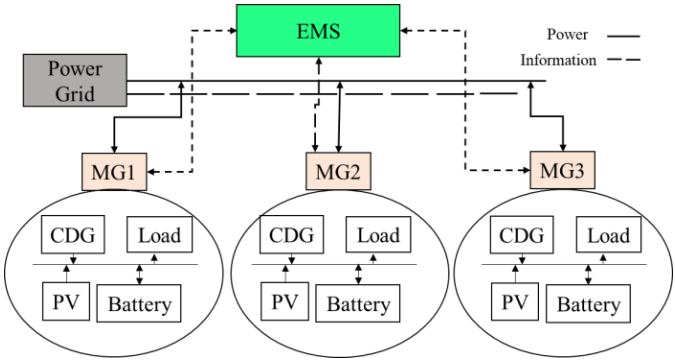

*Fig. 1. The MMG system configuration*

1) Demand Response: It is assumed that loads of MGs consist of controllable and firm loads. Controllable parts of loads are not necessary to be supplied in their own intervals and can be shifted to other time intervals. A price-based DR is implemented in the model to motivate loads to consume their controllable loads in the low-price time intervals. The shortage powers of MGs are meet majorly by taking power from the main grid. Therefore, the DR program is applied based on the prices of purchasing power from the power grid ($\lambda_{(t)}^{buy}$), which fluctuates hourly. Customers are able to gain profit by participating in the DR program and consuming their controllable load in low-price intervals. The gained profit, $\text{Profit}^{DR}$, for each microgrid is calculated according to (1).

$$\text{Profit}^{DR} = P_{(t,t')}^{shift}(\lambda_{(t)}^{buy} - \lambda_{(t')}^{buy}) \quad (1)$$

where $P_{(t,t')}^{shift}$ is the transferred load from $t$ to $t'$, and $\lambda_{(t)}^{buy}$ and $\lambda_{(t')}^{buy}$ are the prices of purchasing power from the power grid at $t$ and $t'$, respectively. When $\lambda_{(t')}^{buy} < \lambda_{(t)}^{buy}$ customers are motivated to shift their controllable load from $t$ to $t'$.

2) Uncertainties: In this study, the uncertainties associated with demand, photovoltaics, and the prices of trading with the main grid are considered to make energy scheduling more practical. The boundaries of load consumption and photovoltaic outputs can be predicted using historical data. To implement the uncertainty of demand and photovoltaic output, the worst case of the predicted data is incorporated in the proposed algorithm. The worst-case occurs when loads consume the upper bounds of the forecasted data, and PVs produce power equal to the lower bounds of the predicted data.

Also, the uncertainties associated with the prices of trading with the main grid are incorporated in the model by the price estimated variance. The risk related to the market prices is formulated as (2):

$$\text{Risk} = \sum_{i=1}^{T}\sum_{j=1}^{T} V_{ij}^{est} \times P_i \times P_j \quad (2)$$

where $V^{est}$ is $T \times T$ matrix, which is determined by the real and predicted prices of $D$ days ahead of the scheduling day according to (3):

$$V^{est} = (1-\alpha)\sum_{i=1}^{D}\alpha^{i-1}(\Lambda_{D-i+1}^{act} - \Lambda_{D-i+1}^{for}) \\ * (\Lambda_{D-i+1}^{act} - \Lambda_{D-i+1}^{for})' \quad (3)$$

where parameter $\alpha$ ($0 \leq \alpha \leq 1$) is used to increase the effect of days nearer to the scheduling day by multiplying greater coefficients to them. To make sure that the covariance matrix is positive definite, it is required that D be at least equal to 24. Notably, the vector $\Lambda$ is $\Lambda = [\lambda_1, \lambda_2, ..., \lambda_{24}]'$.

## IV. COOPERATIVE GAME MODEL FOR MMG SYSTEMS

Several rational players that coordinate together and pool their winning create a cooperative game. Two essential parts of a cooperative game include a set of players and a characteristic function, $v$, which is the value created by different coalitions. A subset of players creates a coalition. A grand coalition is created by all the players [31, 32].

In the MMG system, it is assumed that each microgrid is a player, and all the players form a grand coalition. The characteristic function is assumed as the operational cost. Microgrids transact energy with each other to optimize the cost of the MMG system and, subsequently, their allocated costs. For all the possible coalitions, the allocated cost to each microgrid should be less than their costs in the isolated mode;

otherwise, microgrids will leave the game. In the following, the optimization problem, including the objective function and constraints to achieve the minimum cost for the MMG system, is presented. Then, the total cost is fairly allocated between microgrids by applying Shapley value from cooperative game theory. Notably, the schedule horizon of the MMG system is adjustable and assumed 24-h with hourly time intervals.

*A. Optimization Problem Formulation*

*1) Objective Function*

All the microgrids form a grand coalition, and their objective is shown in (4). The objective function includes the generation cost of CDGs, the cost of transacting power with the power grid, DR, and trading power price risk terms. In (4), the parameter $r$ is considered to adjust the level of price uncertainty in the model. Greater values of $r$ show that the system operator is more risk-averse.

$$\begin{aligned}\text{Min} \sum_t \sum_k \lambda^{cdg}(k,t) * P^{cdg}_{(k,t)} \\ + \sum_t \lambda^{buy}_{(t)} * P^{buy}_{(t)} - \lambda^{sell}_{(t)} * P^{sell}_{(t)} \\ + \sum_k \sum_t P^{shift}_{(k,t,t')} * \left(\lambda^{buy}_{(k,t')} - \lambda^{buy}_{(k,t)}\right) \\ + r * (\sum_t \sum_j V^{est-sell}_{(t,j)} . P^{buy}_{(t)} . P^{buy}_{(j)} \\ + \sum_t \sum_j V^{est-buy}_{(t,j)} . P^{sell}_{(t)} . P^{sell}_{(j)} )\end{aligned} \quad (4)$$

*2) Constraints*

The generation limits of CDGs are shown in (5). Also, the general model of the per-unit cost of CDGs is presented in (6). According to (6), the price of CDGs changes stepwise according to the range of output power. The power balance equation of the system is shown in (7). The binary variable $u(k,t)$ prohibits the charging and discharging of BESS concurrently.

$$P^{min}_{(k)} \leq P^{cdg}_{(k,t)} \leq P^{max}_{(k)} \quad (5)$$

$$\lambda^{cdg}_{(k,t)} = \begin{cases} \lambda_1 & P^{max(0)}_{(k)} \leq P^{CDG}_{(k,t)} \leq P^{max(1)}_{(k)} \\ \lambda_2 & P^{max(1)}_{(k)} < P^{CDG}_{(k,t)} \leq P^{max(2)}_{(k)} \\ \lambda_3 & P^{max(2)}_{(k)} < P^{CDG}_{(k,t)} \leq P^{max}_{(k)} \end{cases} \quad (6)$$

$$\begin{aligned}\sum_k P^{pv}_{(k,t)} + P^{CDG}_{(k,t)} + P^{short}_{(k,t)} + u(k,t) * P^{B-}_{(k,t)} \\ = \sum_k P^{l-adj}_{(k,t)} + P^{sell}_{(k,t)} \\ + (1 - u(k,t)) * P^{B+}_{(k,t)} \quad \forall t \in T\end{aligned} \quad (7)$$

State of Charge (SOC) of BESSs, and their charging and discharging powers are constrained by (8)-(14). Eq. (8)-(11) shows the charging and discharging power limits of BESS. Also, the SOC of BESS is obtained by (12) and bounded by (13) at each interval. Notably, the initial value of SOC at the first interval is set equal to the value of the preceding day's last period. $SOC^{B'}_t$ and $SOC^B_t$ shows SOC of BESS before and after considering self-discharge at $t$, respectively. $SOC^{B'}_t$ and $SOC^B_t$ are related to each other by (14). Back-to-back (BTB) converters are used to connect BESSs to the multi-microgrid system. The losses related to converters are implemented in the suggested model [8].

$$0 \leq P^{B+}_{(k,t)} \leq (1 - u(k,t)) * P^{cap}_B (1 - SOC^B_{(k,t-1)}) \frac{1}{1-L} \\ * \frac{1}{\eta^{BTB}_B} \quad (8) \\ \forall t \in T,, \forall k \in K$$

$$0 \leq P^{B-}_{(k,t)} \leq u(k,t) * P^{cap}_B * SOC^B_{(k,t-1)} * (1 - L^{B-}) \\ * \eta^{BTB}_B \quad (9) \\ \forall t \in T,, \forall k \in K$$

$$0 \leq P^{B+}_{(k,t)} \leq (1 - u(k,t)) * \frac{P^{BTB}_B}{\eta^{BTB}_B} \quad \forall t \in T,, \forall k \in K \quad (10)$$

$$0 \leq P^{B-}_{(k,t)} \leq u(k,t) * \frac{P^{BTB}_B}{\eta^{BTB}_B} \quad \forall t \in T,, \forall k \in K \quad (11)$$

$$\begin{aligned}SOC^B_{(k,t)} = SOC^B_{(k,t-1)} \\ - \frac{1}{P^{Cap}_B} * (\frac{1}{1-L^{B-}} * \frac{1}{\eta^{BTB}_B} * u(k,t) * P^{B-}_{(k,t)} - (1 - \\ u(k,t)) * P^{B+}_{(k,t)} * (1 - L^{B+}) * \eta^{BTB}_B) \quad \forall t \in T,, \forall k \in K\end{aligned} \quad (12)$$

$$0 \leq SOC^B_{(k,t-1)} \leq 1 \quad \forall t \in T,, \forall k \in K \quad (13)$$

$$SOC^B_{(k,t-1)} = (1 - \delta_B) * SOC^{B'}_{(k,t-1)} \quad \forall t \in T, \forall k \in K \quad (14)$$

The amount load that can be shifted to each interval should be bounded to prohibit forming another peak load. Also, the shifted demand from each interval is bounded by the flexible loads of that interval. Eq. (15) and (16) show these constraints. Using (17), loads of MGs after implementing DR are achieved.

$$\sum_{\substack{t' \in T \\ t' \neq t}} P^{shift}_{(k,t,t')} \leq IF^{max}_{(k,t)}, \quad (15)$$

$$\sum_{\substack{t' \in T \\ t' \neq t}} P^{shift}_{(k,t',t)} \leq OF^{max}_{(k,t)} \quad \forall t \in T, \forall k \in K$$

$$OF^{max}_{(k,t)} = P^{L-con}_{(k,t)} \quad \forall t \in T, \forall k \in K \quad (16)$$

$$P_{(k,t)}^{l-adj} = P_{(k,t)}^{l-fix} + P_{(k,t)}^{l-con} + \sum_{\substack{t' \in T \\ t' \neq t}} P_{(k,t',t)}^{shift} - \sum_{\substack{t' \in T \\ t' \neq t}} P_{(k,t,t')}^{shift} \quad \forall t \in T, \forall k \in K \quad (17)$$

*B. A Shapley Value Cost Allocation Method*

In a cooperative game, microgrids coordinate with each other to minimize the total MMG cost. The obtained total operational cost is required to be fairly allocated among microgrids. In this study, a Shapley value cost allocation method is applied to divide the cost of the multi-microgrid system between microgrids fairly. The Shapley value formula is presented in (18) [32, 34]:

$$Cost_{MG_k} = \sum_s \frac{(|s|-1)!\,(n-|s|)!^*}{n!} [V(s) - V(s - \{i\})] \quad (18)$$

where, $Cost_{MG_k}$ is the cooperative cost allocated to microgrid $k$; $s$ represents a sub-coalition including microgrid $k$; $|s|$ is the number of microgrids in sub-coalition $s$; and $n$ is the total number of microgrids. $[V(s) - V(s - \{i\})]$ is the incremental cost of coalition added by microgrid $k$ joining the coalition. Notably, $V(0)$ is equal to zero. For instance, the allocated cost of microgrid 1 is calculated according to (20):

$$Cost_{MG1} = \frac{0!2!}{3!} [V(\{1\}) - V(\{1\} - \{1\})] + \\ \frac{1!1!}{3!} [V(\{1,2\}) - V(\{1,2\} - \{1\})] + \\ \frac{1!1!}{3!} [V(\{1,3\}) - V(\{1,3\} - \{1\})] + \\ \frac{2!0!}{3!} [V(\{1,2,3\}) - V(\{1,2,3\} - \{1\})] \quad (19)$$

The allocated cost of microgrid 2 and 3 are calculated in a similar way.

## V. NUMERICAL SIMULATION

To assess the efficiency of the cooperative method, the presented model is examined on an MMG system containing three MGs. MMG system data are provided in the following. The MINLP optimization problem is implemented through GAMS and MATLAB, and the results are presented and analyzed in the following sections.

*A. Inputs and parameters*

The prices of trade with the main grid are indicated in Fig. 3. The output generation limits of CDGs are provided in Table 1. The parameters of BESSs and converters connected to them are presented in Table 2. The cost functions of CDGs are presented in (20)-(22). The prices of trade with the power grid are obtained from PJM data (June-August 2015). The parameter $r$ is considered equal to 0.001.

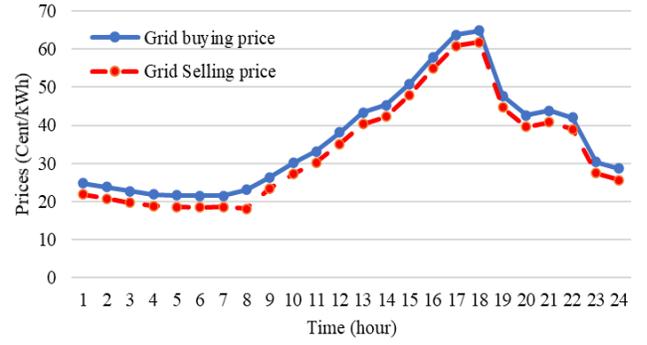

*Fig. 3.* Prices of trading power with the power grid

**Table 1** Output generation limits of CDGs

| Output boundaries | CDGs | | |
|---|---|---|---|
| | MG1 | MG2 | MG3 |
| $P_{min}$ (kWh) | 0 | 0 | 0 |
| $P_{max}$ (kWh) | 500 | 600 | 550 |

**Table 2** BESS and BTB converter parameters [8]

| | |
|---|---|
| Initial charge (kWh) | 50 |
| Storage capacity (kWh) | 200 |
| Discharging loss (%) | 3 |
| Charging loss (%) | 3 |
| BTB convertor capacity (kWh) | 200 |
| BTB convertor efficiency (%) | 98 |

$$\lambda_{(t)}^{cdg(MG1)} = \begin{cases} 25 & 0 \leq P_{(t)}^{cdg} \leq 200 \\ 41 & 200 < P_{(t)}^{cdg} \leq 400 \\ 60 & 400 < P_{(t)}^{cdg} \end{cases} \quad (20)$$

$$\lambda_{(t)}^{cdg(MG2)} = \begin{cases} 25 & 0 \leq P_{(t)}^{cdg} \leq 200 \\ 39 & 200 < P_{(t)}^{cdg} \leq 400 \\ 57 & 400 < P_{(t)}^{cdg} \end{cases} \quad (21)$$

$$\lambda_{(t)}^{cdg(MG3)} = \begin{cases} 16 & 0 \leq P_{(t)}^{cdg} \leq 200 \\ 21 & 200 < P_{(t)}^{cdg} \leq 400 \\ 26 & 400 < P_{(t)}^{cdg} \end{cases} \quad (22)$$

*B. Simulation results*

It is assumed that 6 percent of loads is flexible load and can be shifted to other intervals to gain profit by customers and reduce microgrid cost. Implementing demand response program, load factors of MG1, MG2, and MG3 are enhanced by 4.7%, 5.14%, and 5.2%, respectively.

The output powers of CDGs are presented in Table 3. In the proposed method, the optimization algorithm aims to optimize the cost of the multi-microgrid system. Therefore, CDGs would produce power as far as it is economical for the total system. It is more economical that CDGs with lower per-unit prices generate more power in order to optimize the cost of the muti-microgrid system. It can be seen in Table 3 that the CDG output power of MG3 is set to its maximum capacity and is greater than others; this is due to the lower price of this CDG in comparison with others. Also, MG1, which has the highest per-unit cost, is producing less power than others to optimize the MMG cost.

Considering the uncertainty of trading power prices with the power grid decreases the amount of power transactions with the power grid. Since the uncertainty of trading power prices decreases the trust of units to trade power with the main grid, microgrids would prefer to supply their demands by their CDG generation rather than purchasing power from the power grid. Considering uncertainties, microgrids compromise between supplying their demands with the CDGs and transact energy with the power grid. The buying and selling amount of power with and without considering uncertainty is presented in Table 4.

The charging and discharging schedule of BESS of MG1, MG2, and MG3 are shown in Fig. 4, Fig. 5, and Fig. 6, respectively. BESSs are scheduled to minimize the total MMG cost. Therefore, they are charged in low price intervals and discharged in high price periods to reduce the costs.

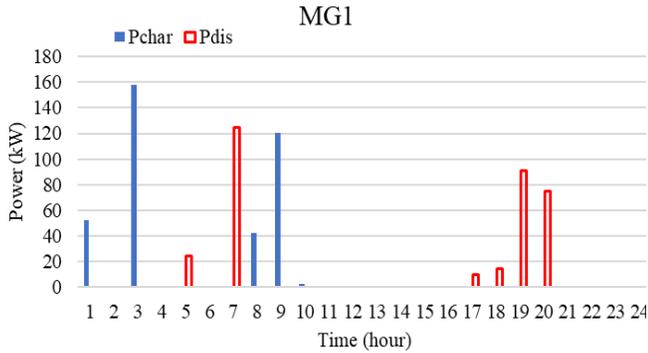

**Fig. 4.** *Performance of BESS of MG1*

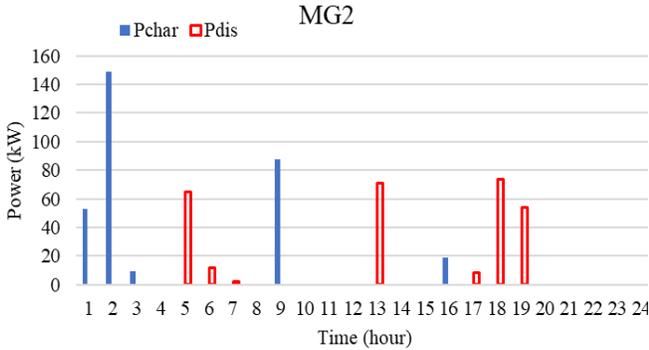

**Fig. 5.** *Performance of BESS of MG2*

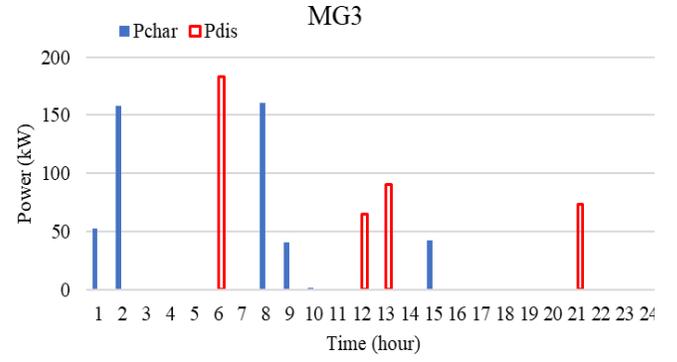

**Fig. 6.** *Performance of BESS of MG3*

It is assumed that each microgrid is a player and all the microgrids form a grand coalition. All the microgrids transact energy with each other to decrease the total cost. To evaluate the effectiveness of the cooperative game to decrease the costs, two cases are evaluated. First, all the microgrids are assumed to be isolated from other microgrids. In this case, microgrids can supply their loads with their own CDGs or trade power with the main grid. Second, all the microgrids form a grand coalition and aim to achieve a global optimum. The cost of the multi-microgrid system for two cases is calculated and provided in Table 5. According to Table 5 the cost of the multi-microgrid system when microgrids participate in the cooperative game is less than the case microgrids are in isolated mode. The cost of the multi-microgrid system is reduced by 4% after forming a coalitional game.

The obtained total operational cost obtained for case II is required to be fairly allocated among microgrids. In this study, a Shapley value cost allocation method is applied to divide the cost of the MMG system between microgrids fairly. The allocated costs of microgrids participating in the cooperative game applying Shapley value are presented in Fig. 7. Also, the cost of microgrids in isolated mode is calculated and presented in Fig. 7. According to Fig. 7, the allocated cost of each microgrid when they participate in the cooperative game is less than their costs in isolated mode. This is due to the fact that microgrids pool their winnings when they participate in the cooperative game and transact with each other to minimize the total MMG cost, and subsequently, their allocated costs. We can see in Fig. 7 that all the microgrids have an incentive to participate in the cooperative game. The results indicate that the costs of MGs participating in the cooperative game are reduced by 2.4%, 2.7%, and 11.8%, respectively.

## VI. CONCLUSION

In this study, a cooperative game approach is presented to schedule the day-ahead operation of multi-microgrid systems. The efficiency of the proposed model is evaluated on an MMG system, including three microgrids. Total cost and each microgrid cost are obtained for two cases. First, all the microgrids are assumed to be isolated from other microgrids. In this case, microgrids can supply their loads with their own CDGs or trade power with the main grid. Second, all the microgrids form a grand coalition and aim to achieve a global optimum. The results indicate that the cost of the multi-microgrid system in the cooperative method is less than the total cost in the isolated status. In the cooperative model, MGs are scheduled to optimize the cost of the multi-microgrid system and achieve a global optimum. Applying Shapley value from the game theory, the cost of the MMG system is fairly allocated between microgrids. The allocated costs of microgrids when they participate in the coalitional game are less than their costs in the isolated status. This is due to the fact that microgrids pool their winnings when they participate in the cooperative game and transact with each other to minimize the total MMG cost and subsequently reduce their allocated costs. Notably, to enhance the confidence level of results, the uncertainty of load, photovoltaic generation, and trading power prices are incorporated in the model. Results show that increasing the degree of price uncertainty decreases customer transactions with the main grid. Also, a price-based demand response is implemented, which motivates loads to consume their flexible loads in low price intervals to reduce operational costs. Applying the demand response program, the load factor of MG1, MG2, and MG3 are improved by 4.7%, 5.14%, and 5.2%, respectively.

**Table 3** Output power of CDGs

|   | MG1 | MG2 | MG3 |
|---|---|---|---|
| t | $P_{cdg}$(kW) | $P_{cdg}$(kW) | $P_{cdg}$(kW) |
| 1 | 200 | 400 | 550 |
| 2 | 351 | 400 | 550 |
| 3 | 400 | 400 | 550 |
| 4 | 200 | 400 | 550 |
| 5 | 400 | 400 | 550 |
| 6 | 400 | 400 | 550 |
| 7 | 400 | 400 | 550 |
| 8 | 400 | 400 | 550 |
| 9 | 400 | 400 | 550 |
| 10 | 400 | 400 | 550 |
| 11 | 400 | 400 | 550 |
| 12 | 400 | 400 | 550 |
| 13 | 400 | 400 | 550 |
| 14 | 400 | 594 | 550 |
| 15 | 400 | 600 | 550 |
| 16 | 400 | 600 | 550 |
| 17 | 400 | 600 | 550 |
| 18 | 400 | 600 | 550 |
| 19 | 400 | 600 | 550 |
| 20 | 400 | 600 | 550 |
| 21 | 400 | 600 | 550 |
| 22 | 400 | 400 | 550 |
| 23 | 200 | 400 | 550 |
| 24 | 247 | 400 | 550 |

**Table 4** Total buying and selling power of MMG system from/to the main grid for $r = 0.001$ and $r = 0$

|   | $r = 0.001$ | | $r = 0$ | |
|---|---|---|---|---|
| t | $P^{sell}_{(t)}$(kW) | $P^{buy}_{(t)}$(kW) | $P^{sell}_{(t)}$(kW) | $P^{buy}_{(t)}$(kW) |
| 1 | 0 | 47.8 | 0 | 797 |
| 2 | 0 | 0 | 0 | 595 |
| 3 | 0 | 0 | 0 | 983 |
| 4 | 0 | 331 | 0 | 1281 |
| 5 | 0 | 120 | 0 | 1359 |
| 6 | 0 | 6.5 | 0 | 1824 |
| 7 | 0 | 0 | 0 | 1277 |
| 8 | 0 | 0 | 0 | 947 |
| 9 | 0 | 171 | 0 | 473 |
| 10 | 0 | 0 | 0 | 546 |
| 11 | 0 | 175 | 0 | 575 |
| 12 | 0 | 144 | 0 | 609 |
| 13 | 0 | 0 | 0 | 561 |
| 14 | 0 | 0 | 0 | 394 |
| 15 | 0 | 0 | 0 | 158 |
| 16 | 0 | 0 | 9 | 0 |
| 17 | 0 | 0 | 0 | 28 |
| 18 | 0 | 0 | 0 | 98 |
| 19 | 0 | 0 | 0 | 345 |
| 20 | 0 | 43.8 | 0 | 520 |
| 21 | 0 | 0 | 0 | 673 |
| 22 | 0 | 108.4 | 0 | 508 |
| 23 | 0 | 99 | 0 | 450 |
| 24 | 0 | 0 | 0 | 397 |

**Table 5** Cost of multi-microgrid system in cooperative game and isolated status

| Mode | Isolated cost ($) | Cost of cooperative game ($) | Cost Saving (%) |
|---|---|---|---|
| MMG system | 1366464 | 1306181 | 4.4 |

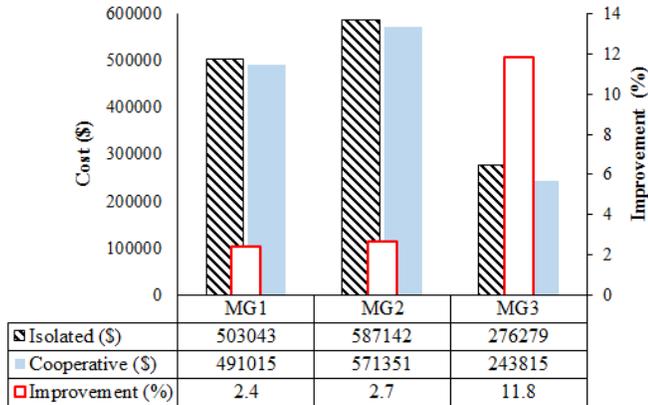

*Fig. 7.* Costs of MGs in cooperative game and isolated status